\documentstyle[amstex,preprint,aps]{revtex}
\title{Viscosity critical behaviour at the gel point in a $3d$ lattice model}
\author{Emanuela Del Gado,$^{a}$ Lucilla de Arcangelis, $^{b}$ and
        Antonio Coniglio $^{a}$}
\address{${}^a$Dipartimento di Scienze Fisiche, Universit\`a di Napoli
         "Federico II",\\ INFM Sezione di Napoli,  Mostra d'Oltremare
         pad. 19, 80125 Napoli, Italy}
\address{${}^b$Dipartimento di Ingegneria dell'Informazione, 
         Seconda Universit\`a di Napoli,\\ INFM Sezione di Napoli, via Roma 29,
 81031 Aversa (Caserta), 
         Italy}
\begin{document}
\maketitle
\begin{abstract}
Within a recently introduced model based on the bond-fluctuation 
dynamics we study the viscoelastic behaviour of a polymer solution at the 
gelation threshold.
We here present the results of the numerical 
simulation of the model 
on a cubic lattice: the percolation transition, the diffusion properties and 
the time autocorrelation functions have been studied.
From both the diffusion coefficients and the relaxation times critical behaviour
a critical exponent $k$ for the viscosity coefficient has been extracted: the 
two results are comparable within the errors giving $k \sim 1.3$, in close 
agreement with the Rouse model prediction and  
with some experimental results. In the 
critical region below the transition threshold the time autocorrelation 
functions show a long time tail which is well fitted by a stretched exponential 
decay.\\
PACS: 05.20.-y, 82.70.Gg, 83.10.Nn
\end{abstract}
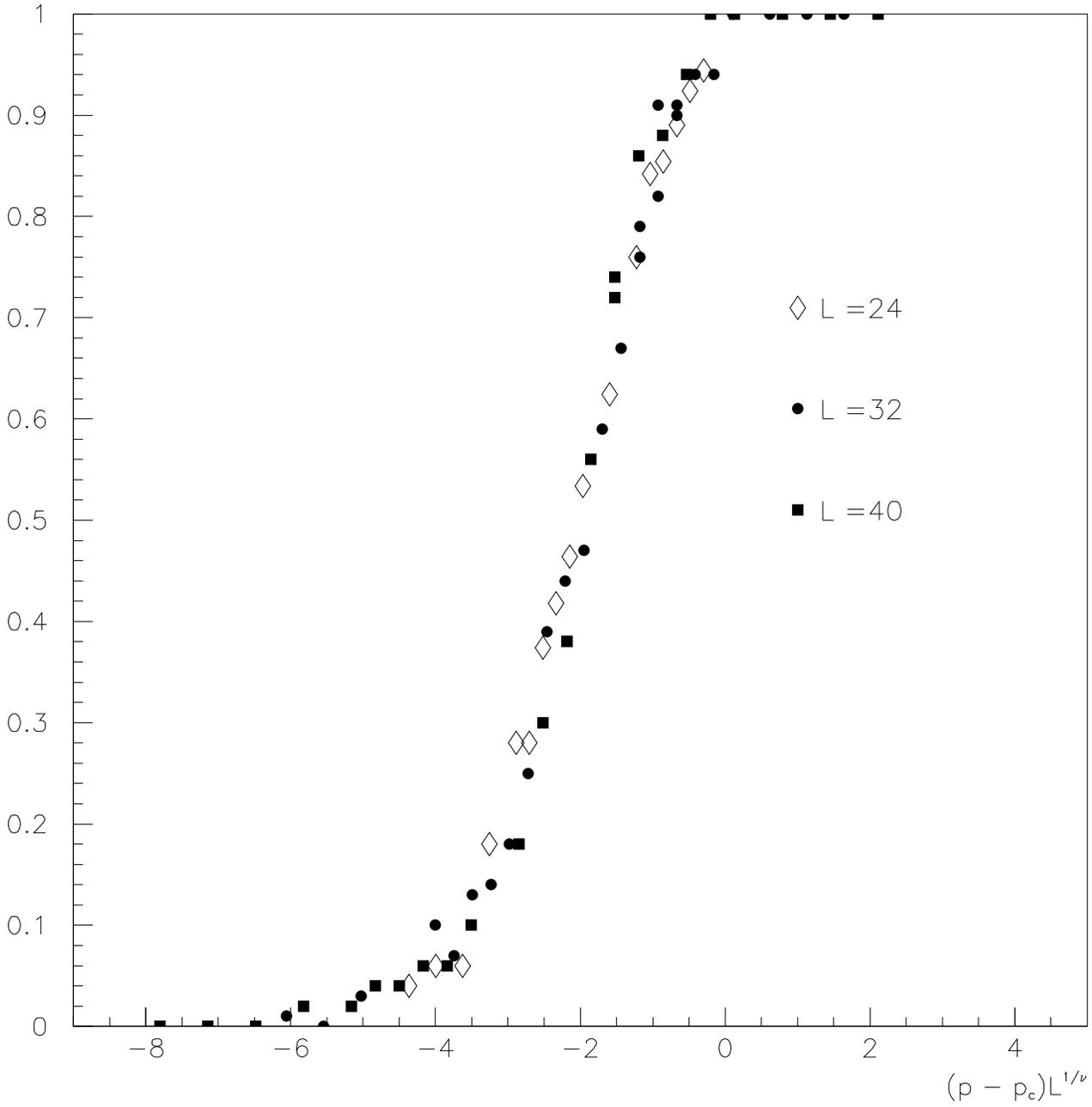
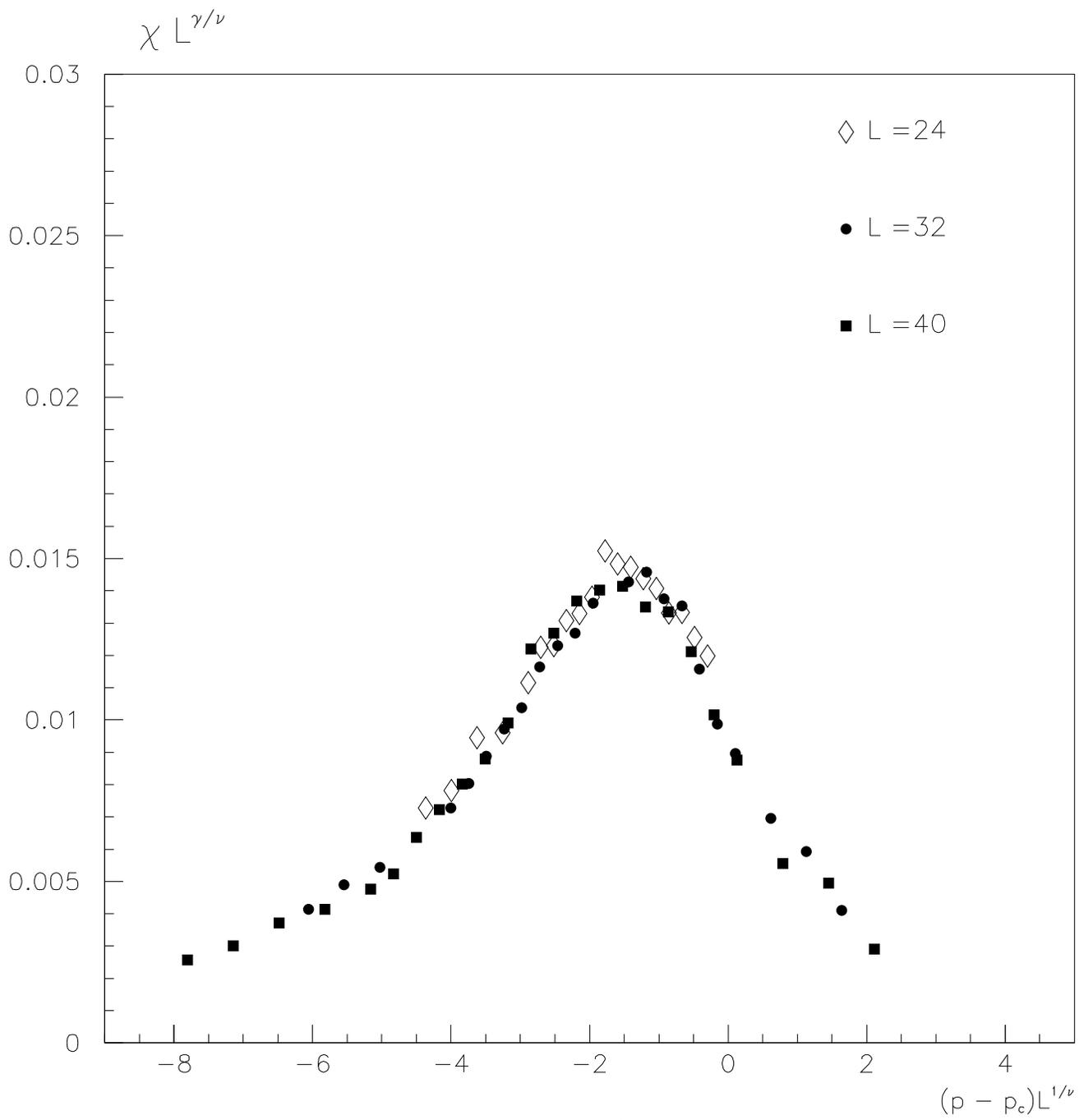
\begin{figure}
\caption{Data collapse for lattice sizes $L=24,32,40$ for the 
percolation probability $P$ $(a)$ and for the mean cluster size 
$\chi$ $(b)$: the best collapse is obtained with 
$p_{c}=0.718 \pm 0.05$, $\nu = 0.89 \pm 0.01$, $\gamma = 1.8 \pm 0.05$.}
\label{fig1}
\end{figure}

%
%
%
\newpage
Polymeric materials are characterized by a rich and complex phenomenology, 
which goes from non newtonian dynamic behaviour in polymer liquids to 
viscoelastic properties of polymer gels and the glass transition of polymer 
melts, intensively investigated in the last decades. The interest 
in such systems has been recently further increased by several 
possibilities of technological applications in many different fields but still 
the non trivial viscoelastic behaviour of polymeric systems does not 
have a completely satisfying description.\\ 
The gelation transition which transforms the polymeric solution, i.e. the sol, 
in a polymeric gel is characterized by a dramatic change in the viscoelastic 
properties: this is usually described in terms of the divergence of the 
viscosity coefficient and the appearance of an elastic modulus which 
characterizes the gel phase \cite{flo,degl}. 
In the experiments both the viscosity coefficient and the elastic modulus 
dependence on the 
polymerization reaction extent are well fitted by a power law 
but the experimental determination of these critical exponents is quite 
controversial: the results are in fact scattered, probably because of the 
practical difficulties in obtaining the gelation transition in a reproducible 
manner, and could only be interpreted on the basis of a better 
comprehension of the relevant mechanisms in the transition. Recent experimental 
measurements of the viscosity critical exponent $k$ give values ranging from 
$0.7$ in diisocyanate/triol to $1.5$ in epoxy resins whereas for the elastic 
modulus critical exponent $t$ the 
values are even more scattered, ranging from $1.9$ in diisocianate/triol to 
$3.0$ in polyesters (see references 
\cite{ad97,ad91,ad92,mawi,col1,col2,nico,malla}).\\
With simple statistical mechanics models it is possible to analyse the 
essential aspects of the transition and its
critical properties. From the Flory model this has led to the description 
of the gelation process in terms of a percolation transition. The percolation 
model has turned out to be the 
satisfactory model for the sol-gel transition \cite{adconst,staul}, it is 
able to describe the role 
of connectivity and gives the critical 
exponents for all the geometrical properties, which are in perfect agreement 
with the experimental results. 
On the other hand the viscoelastic dynamic
behaviour is not simply obtained in terms of the connectivity transition.\\ 
The difficulties in studying the viscosity critical behaviour at the gel point 
come from the determination of the viscosity of a very complex medium, 
which is the sol at 
the transition threshold, a highly polydisperse polymeric solution at high 
concentration. The complex polymer dynamics is characterized by the relaxation 
processes over many different time scales and compete with the increasing 
connectivity to produce the observed viscoelastic behaviour. The simplest 
approach consists in 
considering the sol as a polydisperse 
suspension of 
solid spheres neglecting the cluster-cluster interactions and generalizing the 
Einstein formula for the 
viscosity of a monodisperse suspension of solid spheres 
\cite{adconst}, which corresponds to a highly diluted regime. 
Within the Flory classical theory of gelation the viscosity 
remains finite \cite{flo} or diverges at most logarithmically. 
Using instead the Rouse model for the polymer dynamics \cite{doed}, which neglects 
the entanglement effects and the hydrodynamics interactions, the 
viscosity in a solution of polymeric clusters, expressed 
in terms of the 
macroscopic relaxation time, grows like $<R^{2}>$ as the cluster radius 
$R$ grows in the gelation process. The  
contribution of the $n_{s}$ molecules of size $s$ and gyration radius $R_{s}$ 
to the average $<>$ 
is of the order of $sn_{s}R^{2}_{s}$ leading to the critical exponent 
$k = 2\nu-\beta$ \cite{deg78}, where $\nu$ is the critical exponent 
of the correlation length diverging at the gel point and $\beta$ is 
the critical exponent describing the growth of the gel phase. With the 
random percolation exponents the value $k \sim 1.35$ is found, that 
agrees quite well with the experimental measurements for silica gels of
ref.\cite{mawi,col1}. Actually this Rouse exponent could be 
considered as an upper limit due to 
the complete screening of the hydrodynamic interactions and the entanglement 
effects.
The Zimm approach \cite{doed}, where the monomer correlation due to 
the hydrodynamic interactions are not completely screened, would give a 
smaller exponent.\\
Another approach has been proposed by de Gennes \cite{degl,deg78} using an 
analogy between the viscosity at the gelation 
threshold and the diverging conductivity in the {\em random superconducting 
network} model. Following this analogy an exponent $k \sim 0.7$ is obtained 
in $3d$, according to the determination of the conductivity critical exponent 
in the {\em random superconducting model} \cite{ker}. This result is in 
good agreement with the values experimentally obtained in gelling solutions of 
polystyrene/divinylbenzene and diisocyanate/triol 
\cite{ad91,ad92}.\\ 
Our approach consists in directly investigating the viscoelastic properties at 
the sol-gel transition, introducing within the random percolation 
model the bond fluctuation ($BF$) 
dynamics which is able to take into account the polymer 
conformational changes \cite{carkr1,carkr2}. We study a 
solution of tetrafunctional monomers at concentration $p$ and with a 
probability $p_{b}$ of 
bonds formation. In terms of these two parameters one has different cluster 
size distributions and eventually a percolation transition. Monomers interact 
via excluded volume interactions and can diffuse with local random 
movements. The monomer diffusion process produces a variation of the bond 
vectors  and is constrained by the excluded volume 
interaction and the SAW condition for polymer clusters: these two requirements
can be satisfied if the bond lengths vary within an allowed range. 
This dynamics results to take into account the main 
dynamic features of polymer molecules.\\ 
We have performed numerical simulations of the model on the cubic lattice of 
different lattice sizes ($L=24,32,40$) with periodic boundary conditions. 
The eight sites which are the vertices of a lattice 
elementary cell are simultaneously occupied by a monomer, with the constraint 
that two nearest neighbour ($nn$)  
monomers are always separated by an empty elementary cell, i.e. two 
occupied cells cannot have common sites.
The lattice of cells, with double lattice spacing, has been occupied with 
probability $p$, which coincides with the monomer concentration on the 
main lattice in the thermodynamic limit \cite{dedecon1}. 
Monomers are randomly distributed on the main lattice via a diffusion 
process, then between two $nn$ or next nearest neighbour ($nnn$) monomers bonds 
are instantaneously created with probability $p_{b}$ along lattice directions. 
Since most of the experimental data on the gelation transition refer to 
polymers with monomer functionality $f=4$ we have considered this case 
allowing the formation of at most four bonds per monomer.\\
First, a qualitative phase diagram has been determined, studying the 
onset of the gel phase varying $p$ and $p_{b}$: on this basis we have then 
fixed $p_{b}=1$ and let $p$ vary in the interesting range from the sol to 
the gel phase. 
The percolation transition has thus been studied via the percolation 
probability $P$ and the mean cluster size $\chi$ on lattices of different 
size. From their finite size scaling 
behaviours we have obtained the percolation threshold 
$p_{c} \simeq 0.718 \pm 0.005$, the critical exponents $\nu \simeq 0.89 
\pm 0.01$ for 
the percolation correlation length $\xi$ and 
$\gamma \simeq 1.8 \pm 0.05$ for the mean cluster size $\chi$ 
(fig.(\ref{fig1})). 
These results do agree with the 
random percolation 
critical exponents \cite{staul}. 
\footnote{The same agreement with the random percolation critical exponents has 
already been obtained in the $d=2$ version of the model \cite{dedecon1}. Here 
it is worth to mention that this case on a cubic lattice with monomer 
functionality 
$f=4$ is rather the problem of a {\em restricted valence percolation}. This is 
a percolation on a lattice where the number of bonds emanating from the same 
site is restricted or no site may have more than a fixed number of {\em nn}. 
It does reproduce the occurrence of valence saturation for monomers in the 
gelation process. If the number of allowed bonds per site is greater than $2$  
this problem is expected to belong to the same universality class as 
the random percolation \cite{adconst}. As the restriction on the number of 
bonds per monomer clearly introduces a 
correlation effect in the process of bond formation we have optimized our 
algorithm to minimize this effect and actually obtained a good agreement with 
the random  percolation exponents.}\\ 
The system evolves according to the bond fluctuation dynamics, i.e. the 
monomers diffuse with random local movements along lattice directions within 
the excluded volume constraint and produce the bond length fluctuation among 
the allowed values. In fact the $BF$ dynamics can be easily expressed in terms 
of a lattice algorithm and on a cubic lattice it can be shown that the bond 
lengths which guarantee the SAW condition are 
$l=2,\sqrt{5},\sqrt{6},3,\sqrt{10}$ in lattice spacing units \cite{carkr2}. 
In fig.(\ref{fig2}) a simple example of time evolution of a cluster is 
shown.\\   
We remind that the percolation properties are not modified during the dynamic 
evolution of the system, as it is not possible to break or form bonds, 
which would change the cluster size distribution.\\ 
In order to determine the viscosity critical behaviour we use two independent 
ways, based respectively on the diffusion behaviour of the clusters and 
on the relaxation times. Within the study of diffusion properties in the 
system an interesting picture is obtained with a simple scaling argument on the 
diffusion coefficients as presented in \cite{madwi}: the sol at the sol-gel 
transition is a heteregenous medium formed by the 
solvent and all the other clusters of different sizes, with the mean cluster 
size rapidly growing near the percolation threshold. 
A cluster with gyration radius $R$ can be seen as a probe diffusing in 
such a medium: as long as its radius is much greater than the value of 
the percolation correlation 
length in the sol the Stokes-Einstein relation is expected to be valid, 
and the diffusion 
coefficient $D(R)$ of the probe will decrease proportionally to the inverse 
of the 
viscosity coefficient of the medium, $D(R) \sim 1/R^{d-2} \eta$. Then the 
generic probe of size $R$ diffuses in a medium with a 
viscosity coefficient depending on 
$R$, $\eta(R)$, and a Stokes-Einstein {\em generalized} relation $D(R) 
\simeq 1/R^{d-2} \eta(R)$ should be expected to hold.\\
As the percolation threshold is approached the probe diffuses in a medium 
where a spanning cluster appears with a self-similar structure of 
holes at any length scale. At the 
percolation threshold the viscosity coefficient of the sol (the {\em bulk} 
viscosity coefficient) diverges as $\eta \sim (p_{c} - p) ^{-k}$ and for the 
viscosity coefficient depending on $R$ the scaling behaviour 
$\eta(R) \sim R^{k/\nu}$ should be expected. When $R$ is of the order of the 
correlation length then $\eta(R) \sim \eta$. Following this 
scaling argument $D(R) \sim 1/R^{d-2+ k/\nu}$ at $p = p_{c}$.\\
Within this description the use of the Rouse model would consist in taking 
$D(s)\sim 1/s$ for a cluster of size $s$.
Then for large enough cluster sizes $s$ in percolation $s \sim 
R^{d_{f}}$ where $d_{f}$ is the fractal dimensionality of the percolating 
cluster. Taking $D(R) \sim 1/R^{d_{f}}$ leads to $k\sim (d_{f}+2-d)\nu$, 
which again gives $k = 2\nu -\beta$, the Rouse exponent given in 
\cite{deg78}.\\    
We then study the diffusion of monomers and clusters via the mean square 
displacement of the center of mass. For a cluster of size $s$ (an 
$s$-cluster) this quantity is 
calculated from the coordinates of its center of mass $\vec{R}_{s}(t)$
as
\begin{equation}
 \langle \Delta R^{2}_{s}(t) \rangle = \frac{1}{N_{s}} \sum_{\alpha=1}^{N_{s}}
 (\vec{R}^{\alpha}_{s}(t) - \vec{R}^{\alpha}_{s}(0))^{2}
\end{equation}
where the index alpha refers to the $\alpha$th $s$-cluster and $N_{s}$ is the 
number of $s$-clusters so that this quantity is averaged over 
all the $s$-clusters. All the data refer to a lattice size $L=32$ and the 
calculated  
quantities have been averaged over $\sim 30$ different site and bond 
configurations of the same $(p,p_{b})$ values \cite{dedecon1}.\\
On the basis of the theory of Brownian motion and a simple Rouse model 
approach \cite{doed} the center of mass of a polymeric molecule is expected 
to behave as a 
Brownian particle after a sufficiently long time. Indeed we find a linear 
dependence on time in the long 
time behaviour of $\langle \Delta R^{2}_{s}(t) \rangle$ and determine the 
diffusion coefficient of an $s$-cluster in the environment formed by the 
solvent 
and the other clusters. 
In fig.(\ref{fig3}) it is shown how the asymptotic diffusive behaviour 
is reached after a time which increases with the cluster size $s$, due to the 
more complicated relaxation mechanism linked to the inner degrees of freedom in 
the molecule. As $p$ increases towards $p_{c}$ the $s$-clusters move in a 
medium which is more viscous and whose structure is more and more 
complex. As a consequence, we observe an immediate increase in the time 
necessary to 
reach the asymptotic diffusive behaviour of the center of mass motion for all 
the cluster sizes.\\
At $p=p_{c}$ we have calculated the diffusion coefficients of clusters of 
size $s$ and radius of gyration $R$ in order to obtain the 
dependence $D(R)$. 
The diffusion coefficients decrease with the increasing size of the 
cluster, as it is expected, but after a gradual decrease their values 
dramatically go to zero. This behaviour which is due to the block of 
the diffusion for finite 
cluster size in a finite system has allowed to consider only cluster size up 
to $s=30$. 
In fig.(\ref{fig4}) we have plotted $D(R)$ for different cluster sizes 
($s=5$ to $30$): the data results to be well fitted by a power law behaviour. 
Using this scaling argument which gives the prediction $D(R) \sim 
1/R^{1+k/\nu}$ we obtain for the viscosity coefficient a critical exponent 
$k \sim 1.3 \pm 0.1$.\\
We here briefly mention that on the other side the diffusion coefficients of 
very small clusters are not expected to be linked to the macroscopic 
viscosity, and in fact the diffusion coefficient of 
monomers does not go to zero at $p_{c}$, but has a definitively non-zero 
value for $p > p_{c}$ (fig.(\ref{fig5})). It is also interesting to notice that 
the data seems to agree with a dependence $ e^{-1/1-p}$, suggesting 
some cooperative mechanism in the diffusion process.\\ 
In order to study the viscosity in the system independently from the 
scaling hypothesis given before, we 
have studied the relaxation times via the density time autocorrelation 
functions. We 
have calculated the time autocorrelation function $g(t)$ of the number of 
pairs of $nn$ monomers $\varepsilon(t)$, defined as
\begin{equation}
g(t)= \langle \frac{ \overline{\varepsilon(t')\varepsilon(t'+t)} - \overline{
 \varepsilon(t')}^{2}}
{\overline{ \varepsilon(t')^{2}}  - \overline{\varepsilon(t')}^{2}} \rangle
\label{autc}
\end{equation}
where the bar indicates the average over $t'$ (of the order of $10^3$ time 
intervals) and the brackets indicate
the average over about $30$ different initial site and bond configurations.\\
At different $p$ values in the critical region after a fast transient 
$g(t)$ decays to zero but can not be fitted by a simple time exponential 
behaviour. This is a sign of the existence of a distribution of relaxation 
times which 
cannot be related to a single time and this behaviour had already been 
observed in the $d=2$ study of the model. It is a typical feature of polymeric 
systems where 
the relaxation process always envolves the rearrangement of the system over 
many different length scales \cite{ferry,daoud}. This idea of a 
complex relaxation behaviour is further confirmed by the good fit of the 
long-time decay of the $g(t)$ with a stretched exponential law 
(fig.(\ref{fig6})). This 
behaviour of the relaxation functions is considered typical of complex 
materials and usually interpreted in terms of a very broad distribution of 
relaxation times or eventually an infinite number of them and it is in fact 
experimentally 
observed in a sol in the gelation critical regime \cite{mawio,adprl}. The 
picture we obtain via the density time autocorrelation functions is coherent 
and very close to the experimental characterization of the sol at the gelation 
threshold.\\
For $p \sim p_{c}$ we have then fitted the 
$g(t)$ with a stretched exponential 
behaviour $e^{(-t/\tau_{0})^{\beta}}$ (fig.(\ref{fig7})), 
where $\beta \sim 0.3$. This value is quite lower than the ones experimentally 
obtained in ref.\cite{mawio} for a gelling solution or analitically predicted 
in ref.\cite{chen} for randomly branched polymers: this discrepancy could be 
due to the fact that our data refer to a quite narrow region near the 
gelation threshold where $\beta$ is expected to assume the lowest value. On 
the other hand this value of $\beta$ agrees with the experimental results in 
ref.\cite{adprl} and with the asymptotic value of the stretched exponential 
exponent $\beta$ in both Ising spin glasses and polymer melts close to the 
freezing point in ref.\cite{campb}.\\
The characteristic time $\tau_{0}$ varies with $p$ and increases as 
$p_{c}$ is approached. Plotting $\tau_{0}$ as a function of $(p_{c} - p)$ the 
data can be fitted by a power law dependence, with a critical exponent $\sim 
1.27 \pm\ 0.05$ (fig.(\ref{fig8})), very close to the viscosity critical 
exponent obtained from the diffusion properties. This characteristic time 
extracted from the long time relaxation behaviour diverges at the percolation 
threshold because of the passage between two different 
viscoelastic regimes.\\ 
The most immediate way to characterize the distribution of the relaxation 
times in the system is the average characteristic time defined as
\begin{equation}
\tau(p) = \frac{\int_{0}^{t} t' g(t') dt'}{\int_{0}^{t} g(t') dt'}
\label{time}
\end{equation}
which is a typical macroscopic relaxation time and can be directly linked to 
the viscosity coefficient.\\
Numerically in the eq.(\ref{time}) $t$ has been chosen by the condition 
$g(t') \leq 0.001$ for $t'\geq t$.
This characteristic time grows with $p$ and diverges at
the percolation threshold according to the critical behaviour
\begin{equation}
\tau \propto (p_{c}-p)^{-k}
\label{tau}
\end{equation}
with an exponent $k \simeq 1.3 \pm 0.03$ (fig.\ref{fig9}), 
which gives the critical exponent for the viscosity at the sol-gel 
transition. From an immediate comparison between $k$ and the critical 
exponent for $\tau_{o}$ there seems to be a unique power law characterizing 
the divergence of the relaxation times in the system approaching from below 
the transition 
threshold.\\
For a more detailed description of the distribution of relaxation times at the
sol-gel transition together with a study of the frequency dependence of the 
viscoelastic
properties see references
\cite{ad92,adfr,daoud,mawio}.\\ 
The critical exponent obtained for the viscosity critical behaviour is then 
$k \sim 1.3$ and it well agrees with the value of $k$ previously given, 
although independently obtained. This value is quite close to the one 
experimentally measured in silica gels \cite{col1,madwi} and interestingly 
these systems are characterized by 
a polyfunctional condensation mechanism with tetrafunctional monomers, 
which is actually very similar to the case simulated here.  
This result is also close to the value obtained by recent accurate 
measurements in PDMS \cite{ad97}.\\
It does not agree with the {\em random superconducting network} exponent $k=0.7$\cite{ker} 
according to the de Gennes' analogy, whereas 
it is quite close to the Rouse exponent discussed above. We have already 
mentioned that a Rouse-like description of a polymer solution corresponds to a 
complete screening of the entanglement effects and the hydrodynamic 
interactions and is usually considered not realistic enough. Actually   
the entanglement effects 
could be not so important in the relaxation mechanism in the sol on the 
macroscopic relaxation time scale: due to the fractal structure of the gel 
phase the system is in fact quite fluid, probably 
there is no blocking entanglement yet and such temporary entanglements relaxe
on a smaller time scale, not really affecting the macroscopic relaxation time 
\cite{zip}. Furthermore the screening effect of the hydrodynamic 
interactions in a polymeric solution in the semidiluted regime can be quite 
strong, drastically reducing the range of the interactions so that the Rouse 
model results to be in fact very satisfactory \cite{doed}. This could 
reasonably be the case of the sol at the gelation threshold too, and the 
deviation of the real critical exponent from the Rouse value would turn out 
to be actually very small.\\
The numerical simulations have been performed on the parallel Cray-t3e system 
of CINECA (Bologna,Italy), taking about 15000 hours of CPU time. The authors 
aknowledge partial support from the European TMR Network-Fractals 
c.n.FMRXCT980183 and from the MURST grant (PRIN 97). 
This work was also supported by the INFM Parallel Computing and by the European 
Social Fund.\\  
%

%
\begin{figure}
\caption{Four different possible configurations for a cluster formed by four 
monomers in the time evolution according to the bond fluctuation dynamics. We 
consider the subsequent monomer movements: 
in $a$  
the bond lengths are (starting from the upper center bond and clockwise) 
$l=\protect\sqrt{5},3,3,2$; moving forward the upper left monomer 
we have $b$ 
with $l=2,3,3,\protect\sqrt{5}$; from this configuration moving 
right the other left monomer we have $c$ with $l=2,3,
\protect\sqrt{6},\protect\sqrt{6}$; 
then moving right the front 
monomer on the right leads to $d$ and $l=2,
\protect\sqrt{10},\protect\sqrt{6},\protect\sqrt{6}$.
}
\label{fig2}
\end{figure}
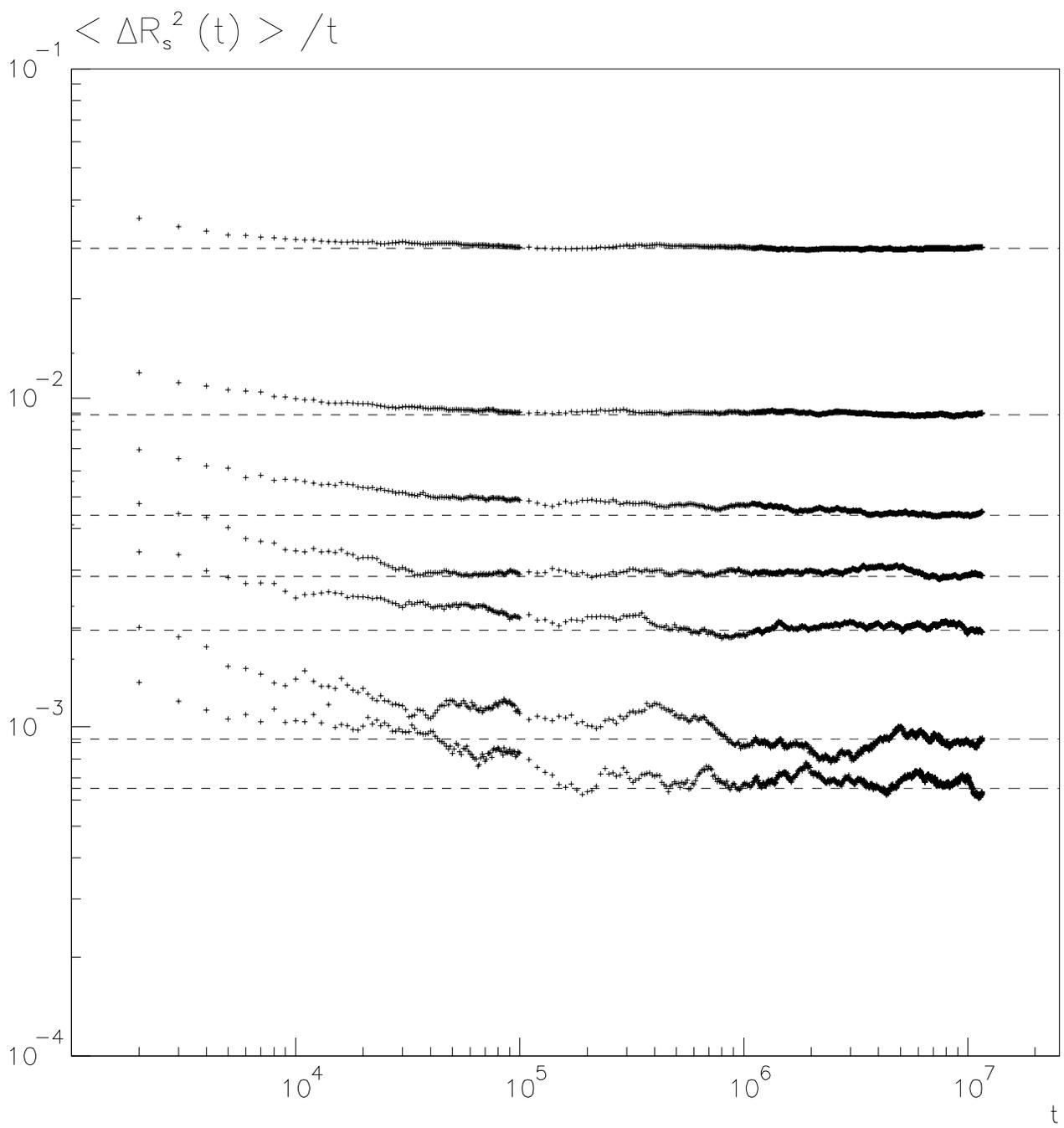
\begin{figure}
\caption{ $\langle \Delta R^{2}_{s}(t) \rangle /t$ as a function of time for 
$p=0.69$ for different cluster sizes; from top to bottom $s=1,2,3,4,5,8,10$. 
The 
broken lines correspond to the asymptotic values, i.e. the values of the 
diffusion coefficients. The data have been averaged over $32$ different 
configurations.}
\label{fig3}
\end{figure}
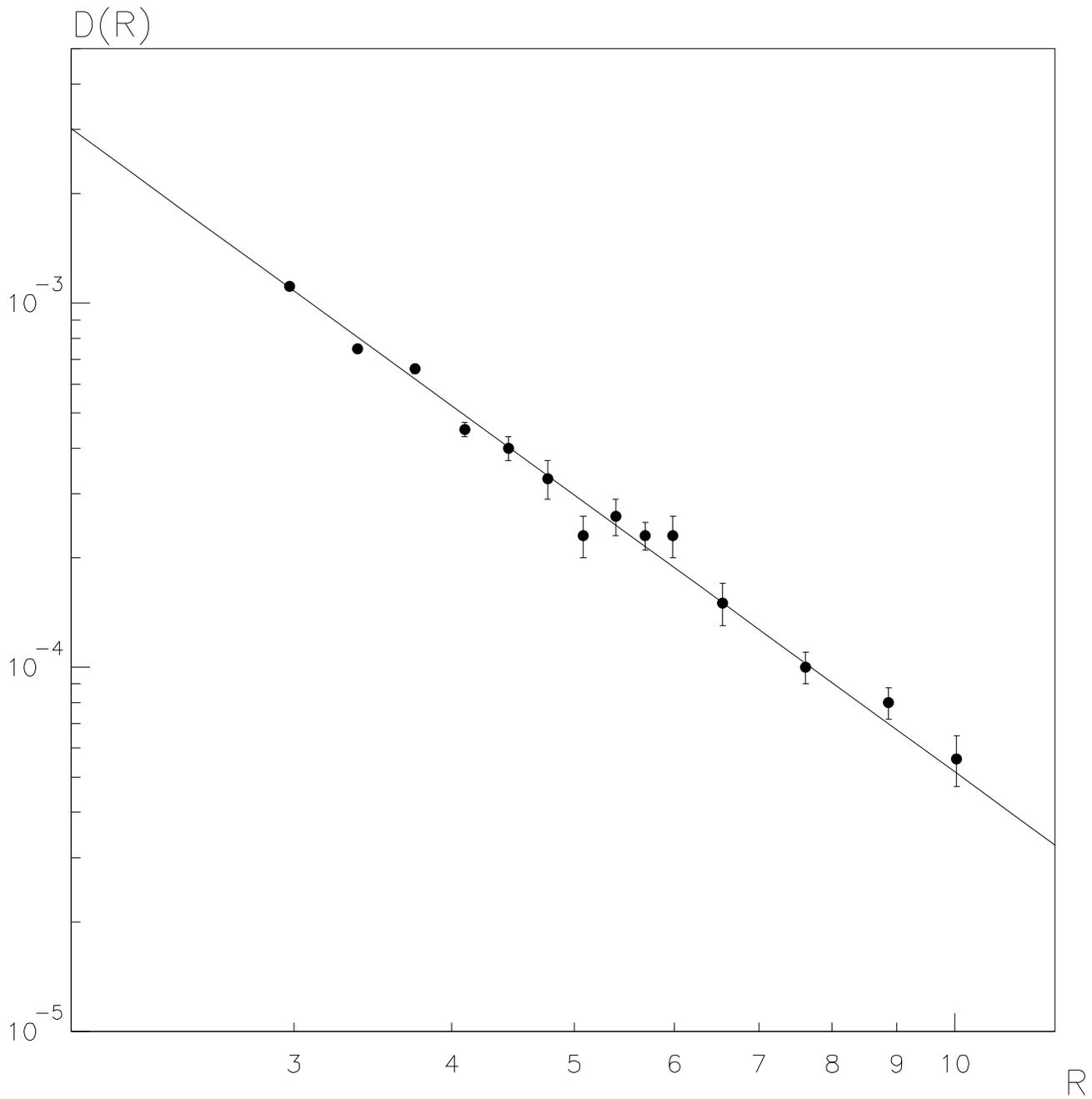
\begin{figure}
\caption{The diffusion coefficient at $p_{c}$ averaged 
over $32$ different configurations for different cluster size as a 
function of the cluster radius of gyration $R$: supposing the scaling 
behaviour $D(R) \sim 1/R^{1+k/\nu}$, gives $k \sim 1.3 \pm0.02$.}
\label{fig4}
\end{figure}
\begin{figure}
\caption{The monomer self-diffusion coefficient as function of $p$. The data 
show how the self-diffusion coefficient of monomers is not linked to the 
macroscopic viscosity diverging at the gel point. $D(s=1)$ becomes 
numerically undistinguishable from zero only at $p > p_{c}$ and the data are 
well fitted with the dependence $\sim e^{-1/1-p}$ (the continuous line).}
\label{fig5}
\end{figure}
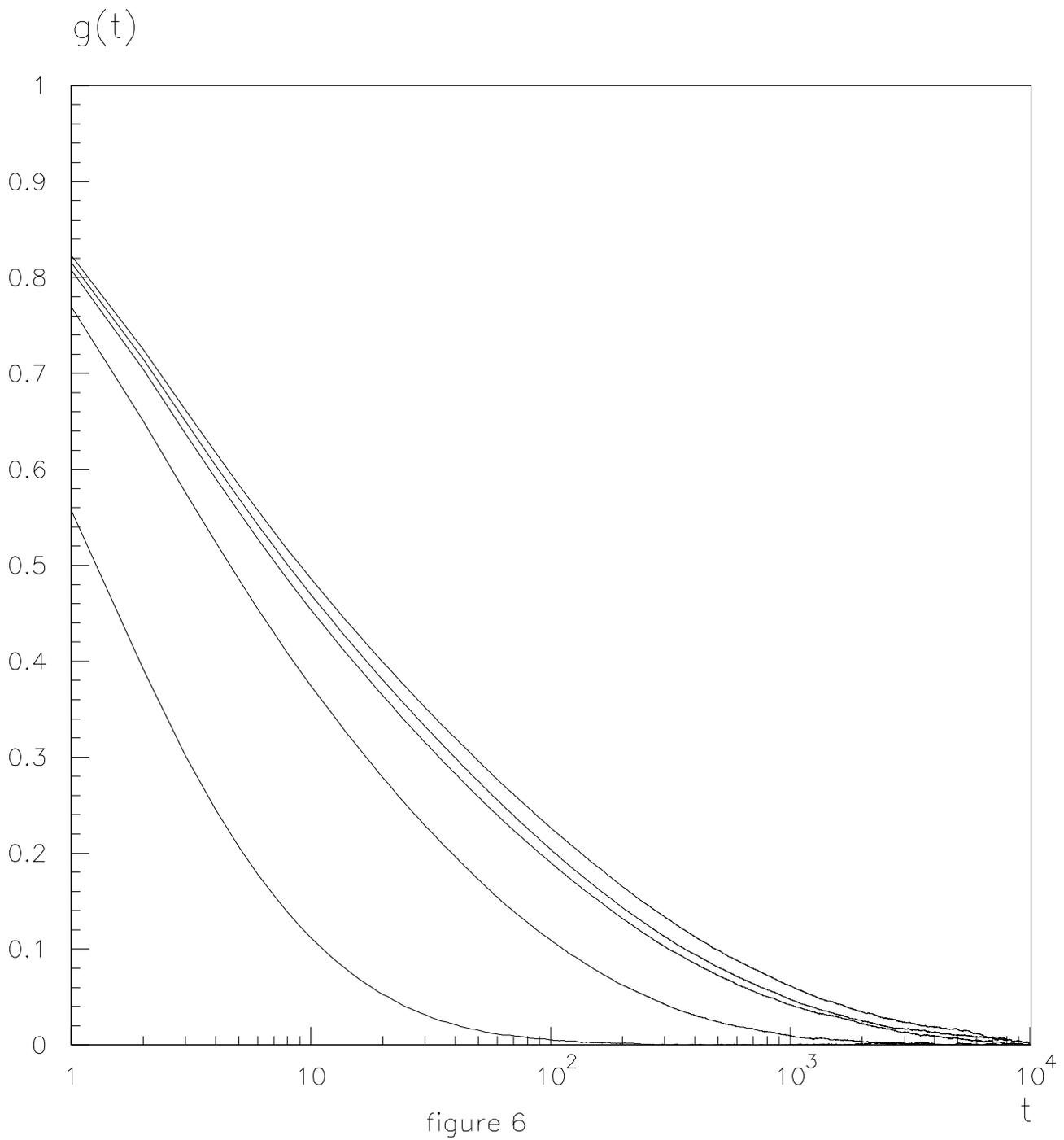
\begin{figure}
\caption{Time autocorrelation function $g(t)$ of the density of $nn$ monomers 
defined in eq.(\ref{autc}) as function of time for various $p$. From top to 
bottom 
$p=0.705, 0.7, 0.69, 0.66, 0.5$ (data averaged over $\sim 30$ different 
configurations).}
\label{fig6}
\end{figure}
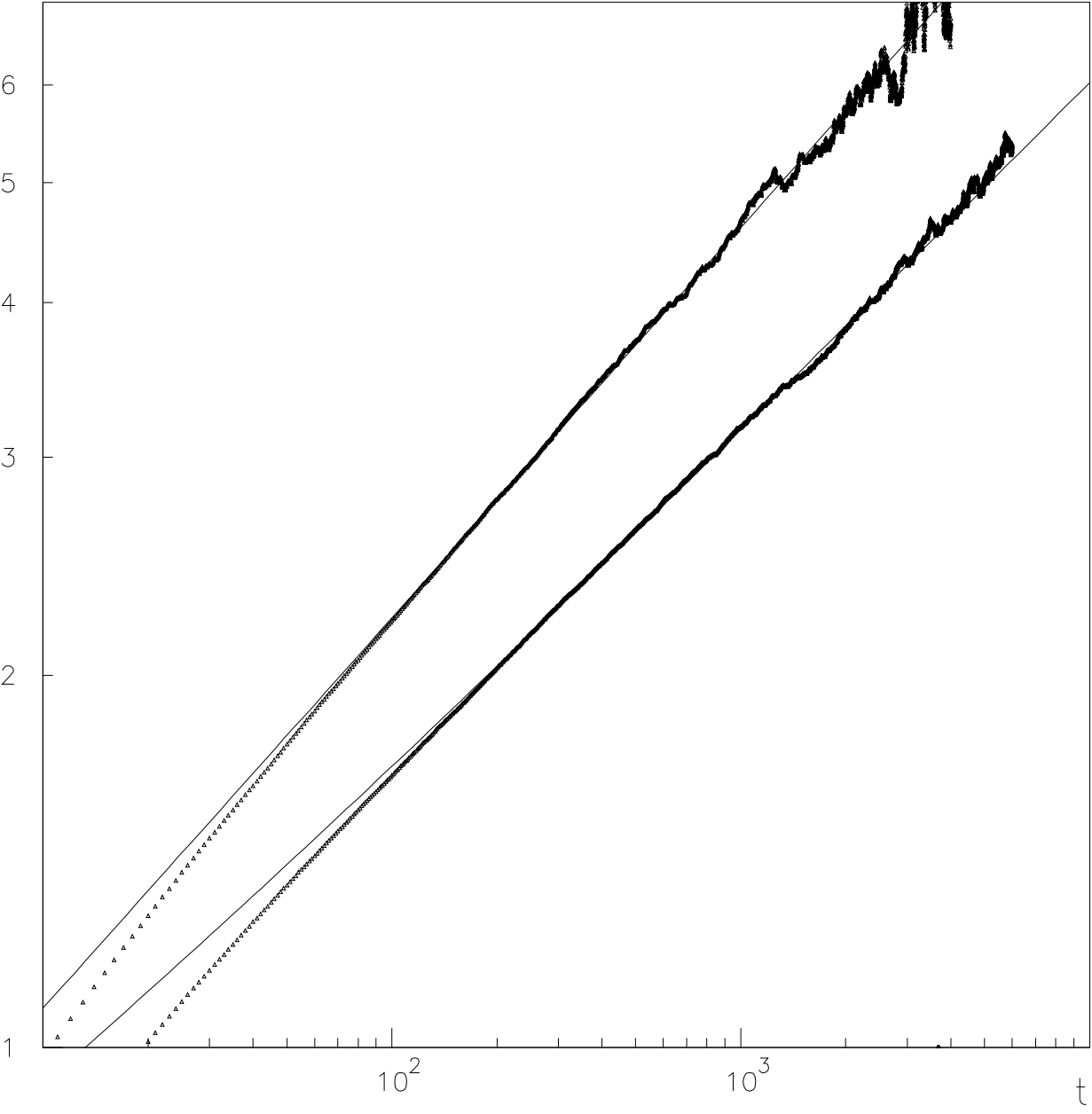
\begin{figure}
\caption{Stretched exponential behaviour $e^{-(t/\tau_{0})\protect ^\beta}$ of 
the long time tail of the time 
autocorrelation function $g(t)$ for $p=0.66$ (bottom) and $p=0.69$ (top).}
\label{fig7}
\end{figure}
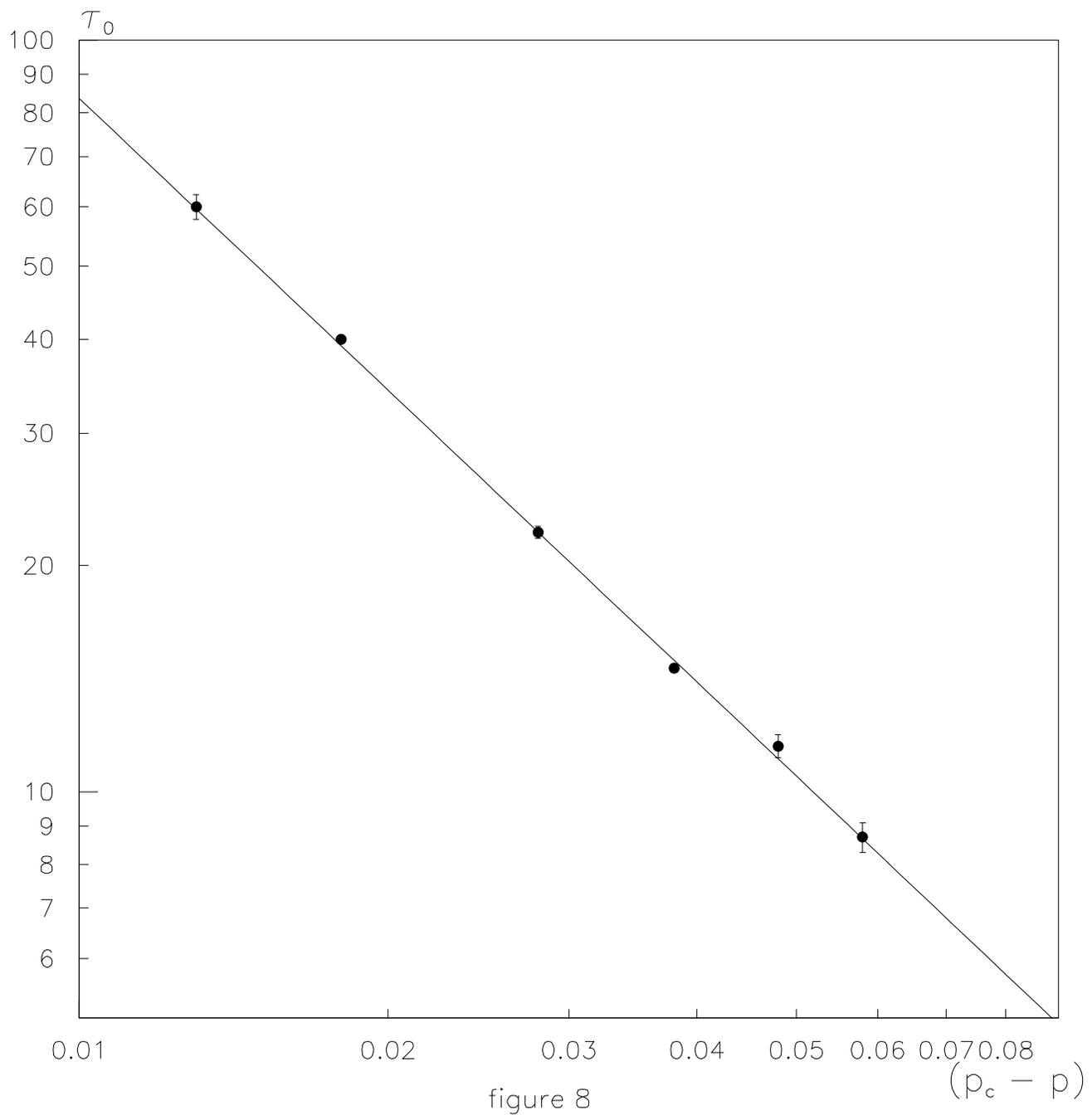
\begin{figure}
\caption{The characteristic time $\tau_{0}$ obtained from the stretched 
exponential fit of the time autocorrelation function long time behaviour as a 
function of $(p_{c} - p)$. The data are well fitted by a power law
with an exponent $1.29 \pm 0.03$.}
\label{fig8}
\end{figure}
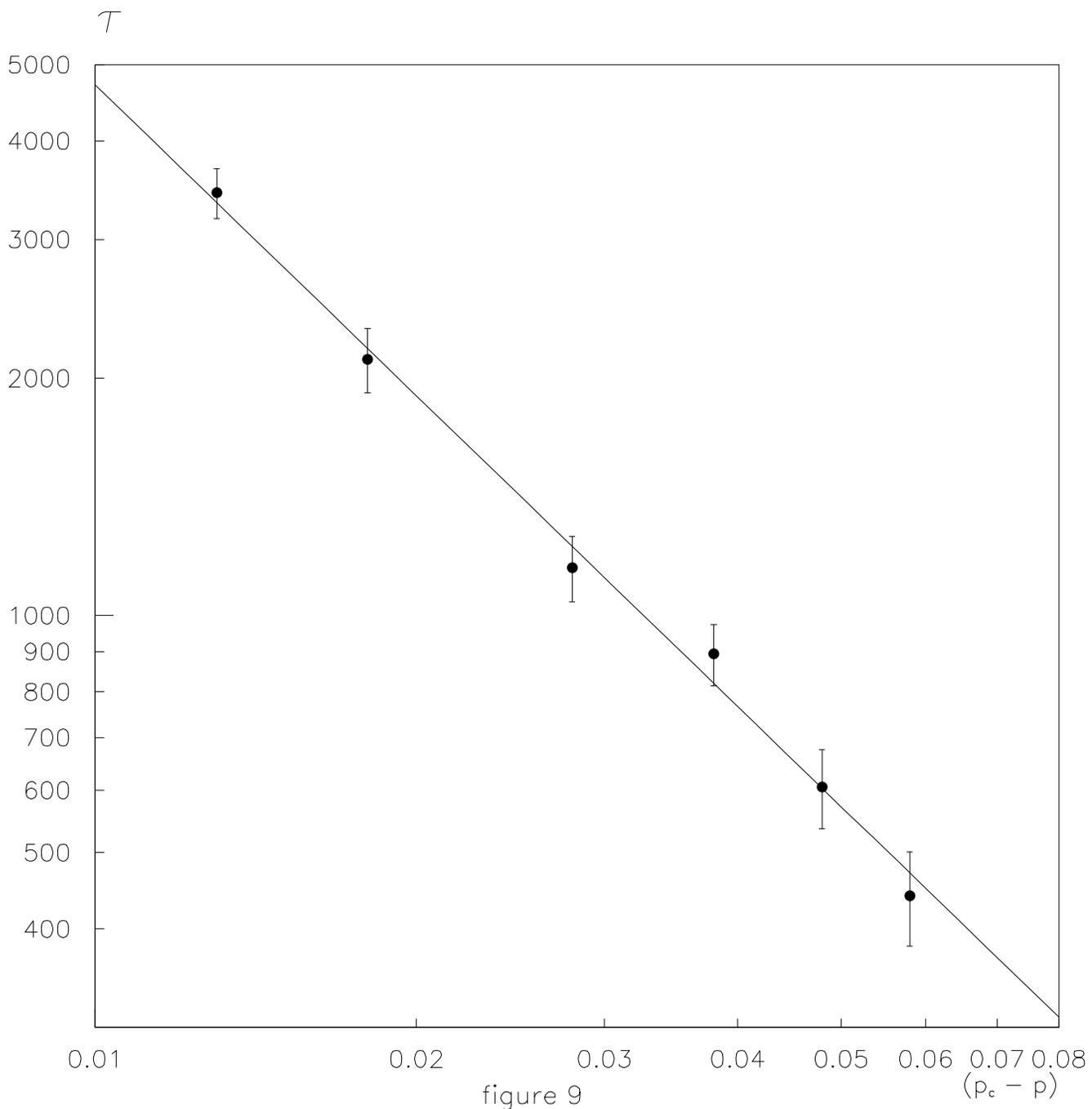
\begin{figure}
\caption{The characteristic integral time $\tau$ calculated according to 
eq.(\ref{time}) as a function of $(p_{c} - p)$. The data are well fitted by a 
power law with a critical exponent $k \sim 1.31 \pm 0.05$.}
\label{fig9}
\end{figure}
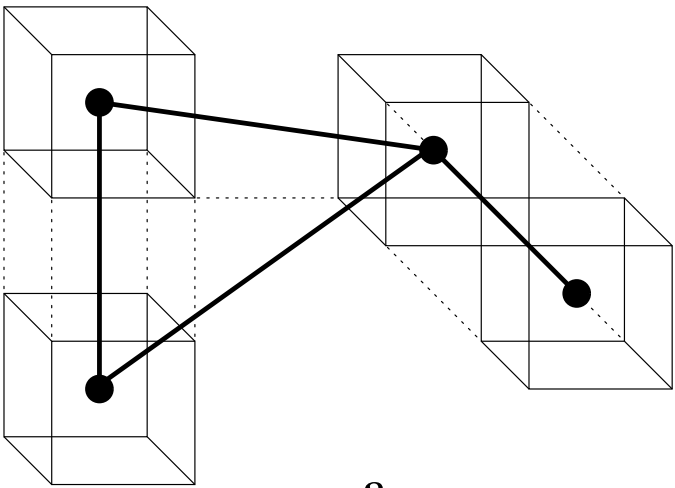
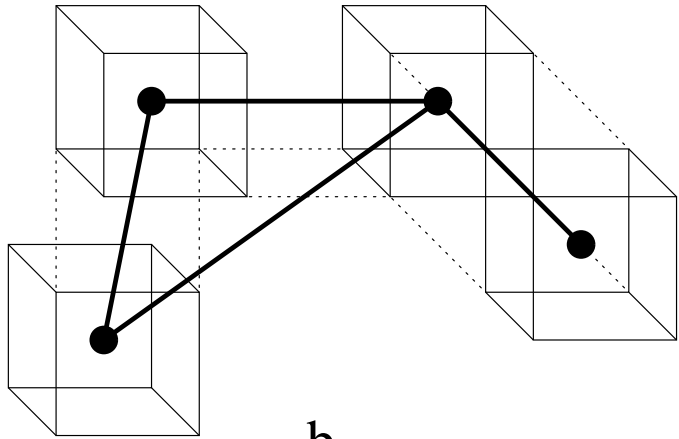
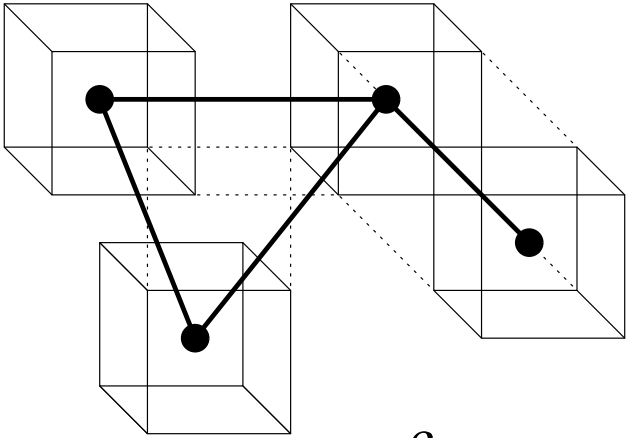
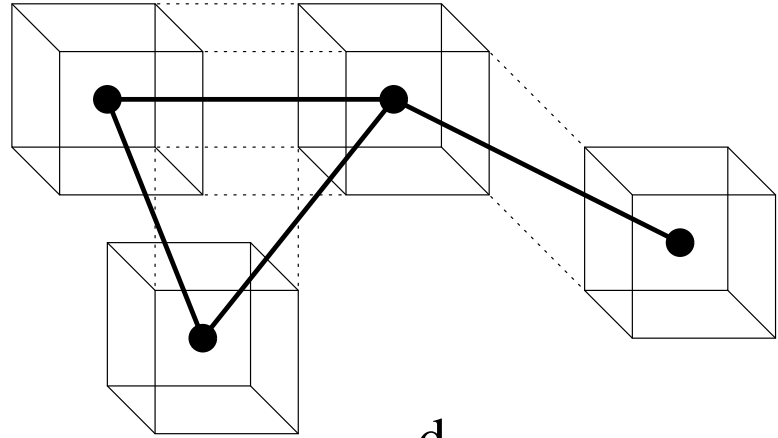
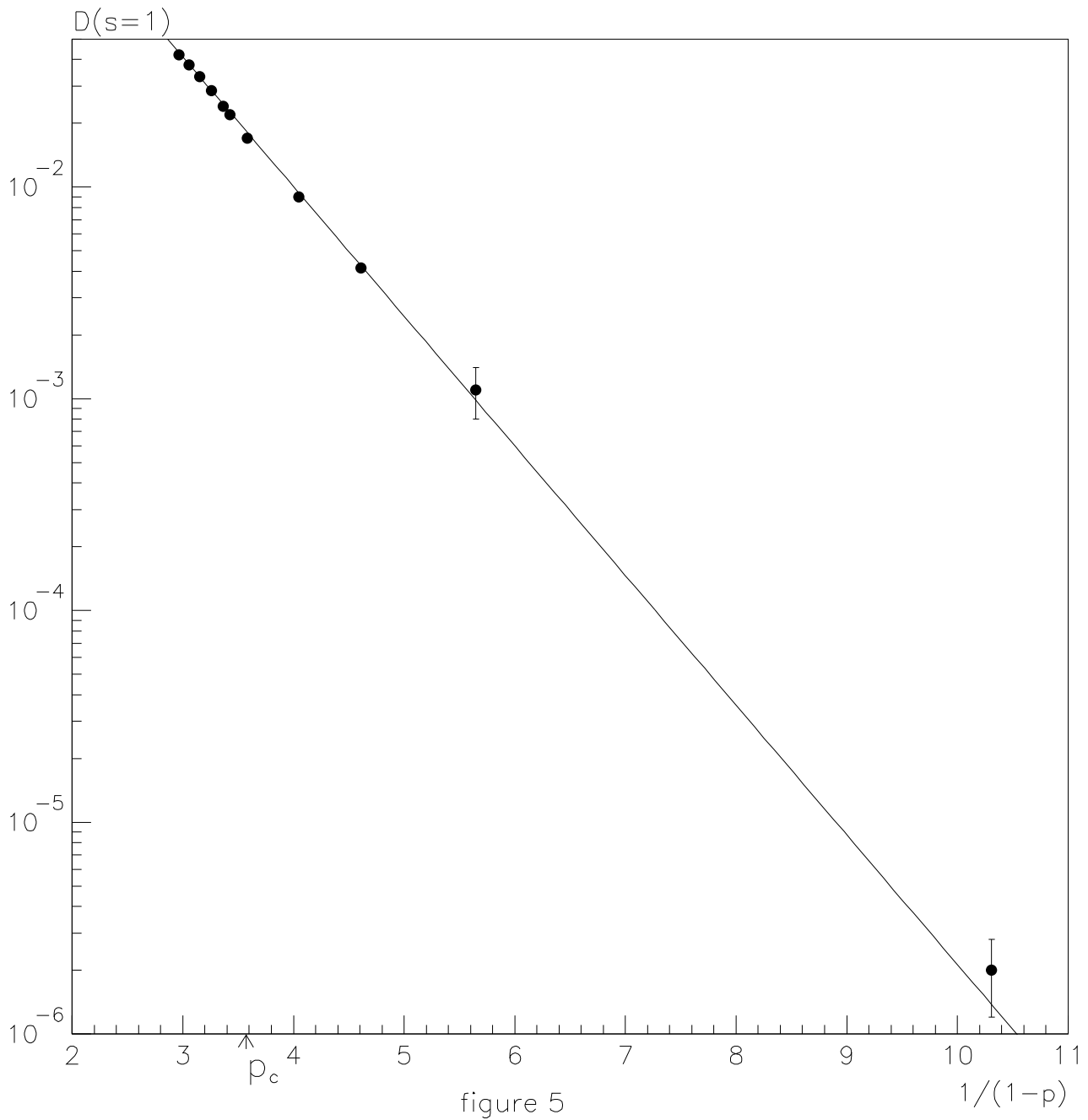
\end{document}